\documentstyle[12pt,openbib,psfig]{article}
\newcommand{\dirac}[1]{#1\hspace{-0.2cm}/\ }
\newcommand{\f}{\frac}

\newcommand{\de}{\delta}

\newcommand{\g}{\gamma}

\newcommand{\ep}{\epsilon}

\begin{document}
\title{\bf Density dependent strong coupling constant of QCD derived 
from compact star data }
\author{ Subharthi Ray $^{1,2,3}$, Jishnu Dey $^{1,2,\dagger}$ and  Mira Dey
$^{3,\dagger}$}
\date{\today }
\maketitle
\begin{center}
{\bf Abstract}
\end{center}

	The present work is an endeavour to connect the properties of tiny
nearly massless objects with those of some of the most massive ones, the
compact stars.

    Since 1996 there is major influx of X-ray and $\g$ ray  data from binary 
stars, one or both of which are compact objects that are difficult to explain 
as neutron stars since they contain a mass M in too small a radius R . The 
suggestion has been put forward that these are strange quark stars (SS) 
explainable in a simple model with chiral symmetry restoration (CSR) for 
the quarks and the M, R and other properties like QPOs (quasi periodic 
oscillations) in their X-ray power spectrum. 

	It would be nice if this astrophysical data could shed some light on 
fundamental properties of quarks obeying QCD. One can relate the strong 
coupling constant of QCD, $\alpha_s$ to the quark mass through the
Dyson-Schwinger gap equation using the real time formalism of Dolan and
Jackiw. This enables us to obtain the density dependence of $\alpha_s$ from
the simple CSR referred to above. This way fundamental physics, difficult to
extract from other models like for example lattice QCD, can be constrained
from present~-~day compact star data and may be put back to modelling the
dense quark phase of early universe.
\vskip .3cm
\noindent (1) IUCAA, Ganeshkhind, Pune 411 007, India\\ 
(2) Azad Physics Centre, Dept. of Physics, Maulana Azad College, Calcutta 700
013, India\\
(3) Dept. of Physics, Presidency College, Calcutta 700
073, India\\ 
\vskip .2cm
$\dagger$ permanent address; 1/10 Prince Golam Md. Road, Calcutta 700 026,
India, IUCAA Senior Associate, e-mail~:~deyjm@giascl01.vsnl.net.in

** Suported by DST Grant no. SP/S2/K18/96, Govt. of India.

\newpage
 	A calculation was done for cold stars in \cite{1} and important
conclusions drawn from there about chiral symmetry restoration in QCD when
the EOS was used to get SS fitting definite mass-radius (M-R) relations \cite
{1, 2, 3}. The empirical M-R relations were derived from astrophysical
observations like luminosity variation and some properties of QPO-s in the
X-ray power spectrum of these compact stars. The calculations are compared to
these stars which emits the X-rays, generated from accretion from their
binary partner.  The different QPO-s show a correlation for not only these
stars but black holes also so that one cannot explain these as being due to
magnetic fields or other properties of the stars but some scale-dependent
phenomenon in the orbits of the particles which are accreted \cite{3}. One of
the compact objects, the SAX J1808.4~-~3658 with period 2.49 millisecond, has
been called the holy grail of X-ray astronomy \cite{vdK}.  Its discovery was
anticipated for nearly 20 years because magnetospheric disk accretion theory
as well as evolutionary ideas concerning the genesis of millisecond radio
pulsars strongly suggested that such rapid spin frequencies must occur in
accreting magnetic field neutron stars.

    The interesting point made by Dey et al. \cite{1} is that starting from
an empirical form for the density dependent masses of the up (u), down (d)
and strange (s) quarks (q in short) given below, one can constrain the
parameter of the form of this mass from  recent astronomical data{\footnote{
$\rho_B = (\rho_u+\rho_d+ \rho_s)/3$ is the baryon number density, $\rho_0 =
0.17~fm^{-3}$ is the normal nuclear matter density, and $\nu$ is a numerical
parameter.  The current quark masses $m_i$ used in the following are 4, 7 and
150 MeV for u, d and s respectively.}}:
\begin{equation}
M_i = m_i + M_Q~~sech(\nu \f{\rho_B}{\rho _0}), \;\;~~~ i = u, d,
s. \label{eq:qm}
\end{equation}

\begin{figure} [here]
\hbox{\hspace{4.5em}
\hbox{\psfig{figure=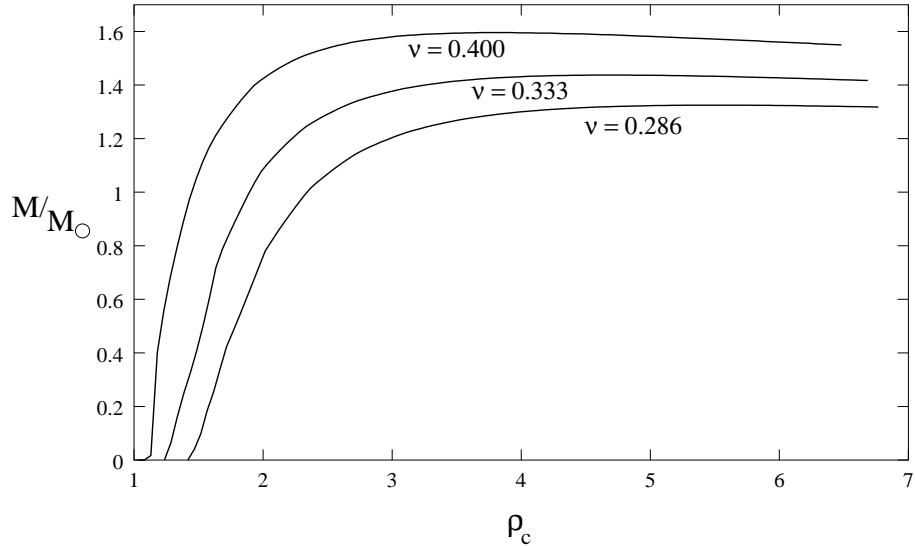,width=12cm}}}
\caption{Variations of the mass with central densities $\rho_c$ (in units of
10$^{15}$ g/cm$^3$), for three different values of the parameter $\nu$ (=
0.286, 0.333 and 0.400) as found in \cite{1}.}
\label{fig:rm}
\end{figure}

	For three values of $\nu~ = ~0.286, ~0.333 $ and $0.400$ the q-star
can have a sequence of masses obtained from the standard TOV equations for
different choices of the central density. This is shown in Fig.
(\ref{fig:rm}).

\vskip 1cm

In other words the masses of stars in units of solar mass,
($M/M_{\odot}$), found as a function of the star radius R, calculated using
the above eqn.(\ref{eq:qm}),  - produces constraints which enable us to
restrict the parameter $\nu$. At high $\rho_B$ the q- mass $M_i$ falls
from its constituent value $M_Q$ to its current one $m_i$. The other
parameter $M_Q$ was taken to be 0.31 $GeV$ to match up with constituent quark
masses assuming the known fact that the hadrons have very little potential
energy. The results are not very sensitive in so far as changing $M_Q$ to
0.32 $GeV$ changes the maximum mass of the star from 1.43735 M$_\odot$ to
1.43738 M$_\odot$ and the corresponding radius changes from 7.0553 kms to
7.0558 kms.

\begin{figure} [here]
\hbox{\hspace{1em}
\hbox{\psfig{figure=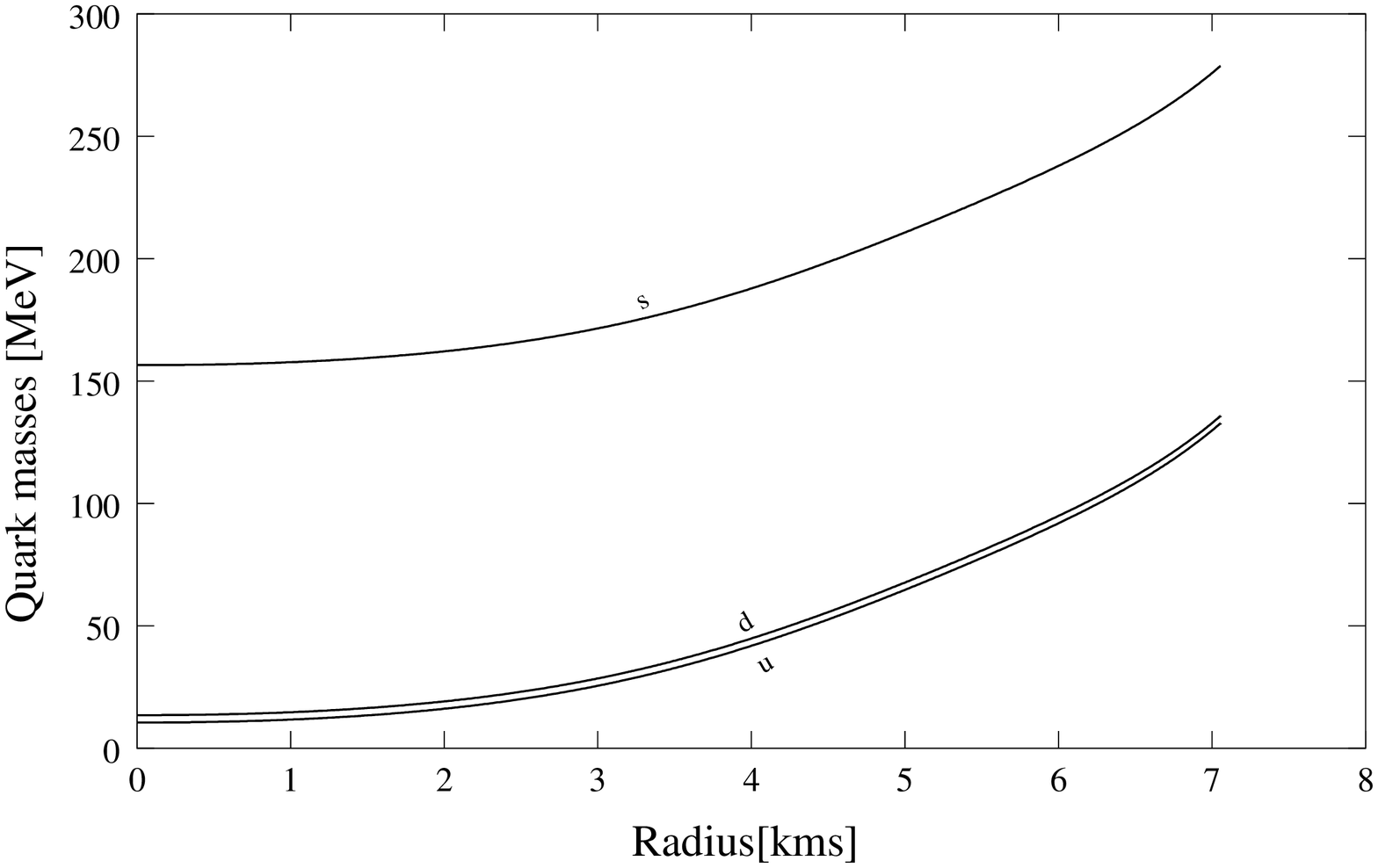,height=8cm}}}
\caption{Quark masses at different radii in the star. Notice that chiral
symmetry is restored inside a radius of 2 kms but outside of this the
masses are substantial.}
\label{fig:chrl}
\end{figure}

    It is interesting to plot the up (u), down (d) and strange (s) quark
masses at various radii in a star. This is done for a particular value of the 
parameter  $\nu = 1/3$ already discussed in \cite{1,2,3} which gives a
sequence of stars falling right into the  allowed region of the M-R curve.
Fig (\ref{fig:chrl}) shows that the quarks do not have the constant current
masses assumed in the bag model nor do they have the constituent masses of
zero density hadrons. Upto a radius about 2 kms the quarks have their chiral
mass but in the major portion of the star their masses are substantially
higher. At the surface the strange q- mass is about 0.278 $GeV$ and the u,d
q- masses $\sim 0.13$ $GeV$.

	Given the form for the mass eq. (\ref{eq:qm}) we have to look for a
formalism to calculate this mass in QCD. This can be done very conveniently
using the real time formalism of Dolan and Jackiw \cite{dj} (DJ in short)
since here one does not encounter the problem of imaginary chemical
potential. As is well known usually people work in the Euclidean space for
lattice and other gauge invariant theories. But this excludes finite density
calculations since that involves imaginary chemical potentials that makes the
action unbounded from below. Using DJ, the price one has to pay is that this
is a gauge dependent formalism. We work in the well established Landau gauge
used by \cite{bcs}. We believe in the physical nature of our results and wish
to impress upon the reader the robustness of this physicality. In other words
we argue that the modelling that we have to invoke does not take away the
basic nature of the results that we obtain, namely that the strong coupling
constant $\alpha_s$ decreases with increasing density and in principle this
can be constrained by compact star data if one believes them to be SS.

\begin{figure} [here]
\hbox{\hspace{1em}
\hbox{\psfig{figure=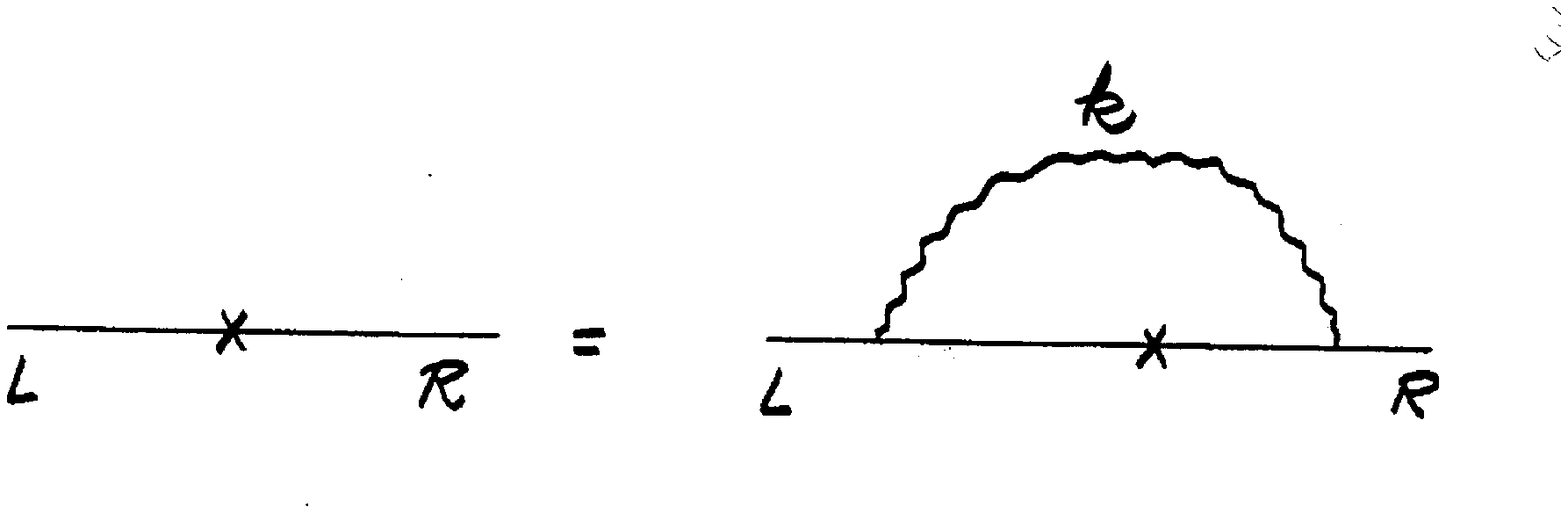,width=14cm}}}
\caption{The self energy of quarks with a gluon loop}
\label{fig:bln}
\end{figure}

The Fermion propagator in the DJ formalism is as follows :
\begin {equation}
S_F(p) = \f{i(\dirac{p} + m)}{p^2 - m^2 + i\ep} - \f{2 \pi\de (p^2 -
m^2)}{exp[{{p_0 -
\ep_F}/T}] + 1}
\label{eq:H}
\end{equation}
where $T$ is the temperature and $\ep _F$, the chemical potential. There is a
similar propagator for the gluon involving the Bose function instead of Fermi
but with no chemical potential.  One has to calculate the self energy of the
quark with this propagator including a gluon loop of four momentum $k$ in
Fig(\ref{fig:bln}).

Using $\g^\mu D_{\mu\nu}\g^\nu ~=~ -3i/k^2$ and a colour factor 4/3 for the 
colour group SU(3) one gets \cite{bcs}   
\begin{equation}
-m(\rho)~ = ~ 4 g_s^2 \f{m(\rho)}{(2\pi)^4} \int d^4 k [\f{i}{k^2[(p-k)^2 - 
m(\rho)^2]} - 
\f{2\pi \de[(p-k)^2 - m(\rho)^2}{k^2} f(p_0 - k_0)]
\label{eq:V}
\end{equation}
where we have neglected a Boson finite $T$-term which deos not contribute and
$f$ is the Fermi function
\begin{equation}
f(E) = \f{1}{1 + exp[(E - E_F)/T]}.
\label{eq:gm}
\end{equation}
In the rest frame, $\vec{p} = 0 $. The Fermi function becomes a step function
for our case, since the temperature T = 0. On closing the contours in the
lower half $k_0$ plane, first term reduces to
\begin{equation}
\int\f{d^3k}{(2\pi)^3}[\f{1}{2E[(m - E)^2 - {\vec k}^2]}-\f{1}{4m{\vec k}^2}]
= \f{4g_s^2 m }{8\pi^2}\int \f{dk}{E}
\label{eq:fir}
\end{equation}
while the second term becomes, on using the variable $p - k \rightarrow q$, 

\begin{equation}
4g_s^2 m\int\f{d^4q}{(2\pi)^2} \f{\de(q^2 - m(\rho)^2)}{(p-q)^2. 
(exp[{(q_0- \mu)\beta}]+1} = \f{-4g_s^2 m}{8\pi^2}\int\f{dq}{E}\f{2}
{exp[{(q_0- E_F)\beta}]+1}
\label{eq:i}
\end{equation}

\begin{table}[here]
\begin{center}
\begin{tabular}{|c|c|c|c|c|}
\hline
$\rho_B/\rho_0$ &$\nu ~=~ $& 0.333&0.286&0.400\\
\hline
\hline
       5 & &  0.5093  &  0.522 &  0.5156\\ \hline
       6 &  & 0.5077 &  0.5025  &  0.478\\ \hline
       7  & & 0.4919 &  0.4739 &  0.4384\\ \hline
       8  & & 0.4693 &  0.4427 &  0.4012\\ \hline
       9 &  &  0.444 &   0.412 &  0.3676\\ \hline
      10 & &  0.4185 &  0.3833 &  0.3379\\ \hline
      11 & &  0.3939 &  0.3571 &  0.3119\\ \hline
      12 & &  0.3707 &  0.3334 &  0.2891\\ \hline
      13 & &  0.3493 &  0.3121 &  0.2691\\ \hline
      14 & &  0.3297 &  0.2929 &  0.2516\\ \hline
      15 & &  0.3117 &  0.2758 &  0.2361\\ \hline
      16 & &  0.2953 &  0.2604 &  0.2223\\ \hline
      17 & &  0.2803 &  0.2465  &   0.21\\ \hline
      18 & &  0.2666 &  0.2339  &  0.199\\ \hline
\hline
\end{tabular}
\end{center}
\vskip 1cm
\caption{Variation of $\alpha_s$ with density for three values of $\nu$}
\end{table}

The first integral is logarithmically divergent and therefore must be set to 
renormalize the $\rho = 0, ~ T \equiv 1/beta ~= 0$ quark mass. We do not use
a high value of $g_s^2/4\pi~ \equiv ~\alpha_s = 0.75 $ like \cite{bcs} since
it gives a cut-off which is too small, of magnitude $1.29~ GeV$ only.  We
prefer a lower $\alpha_s = 0.55$ and this also gives us a high cut-off $k_D
~= ~ 2.86~ GeV$.  We use $m(\rho = 0) = 0.33 ~GeV.$

\begin{figure} [here]
\hbox{\hspace{1em}
\hbox{\psfig{figure=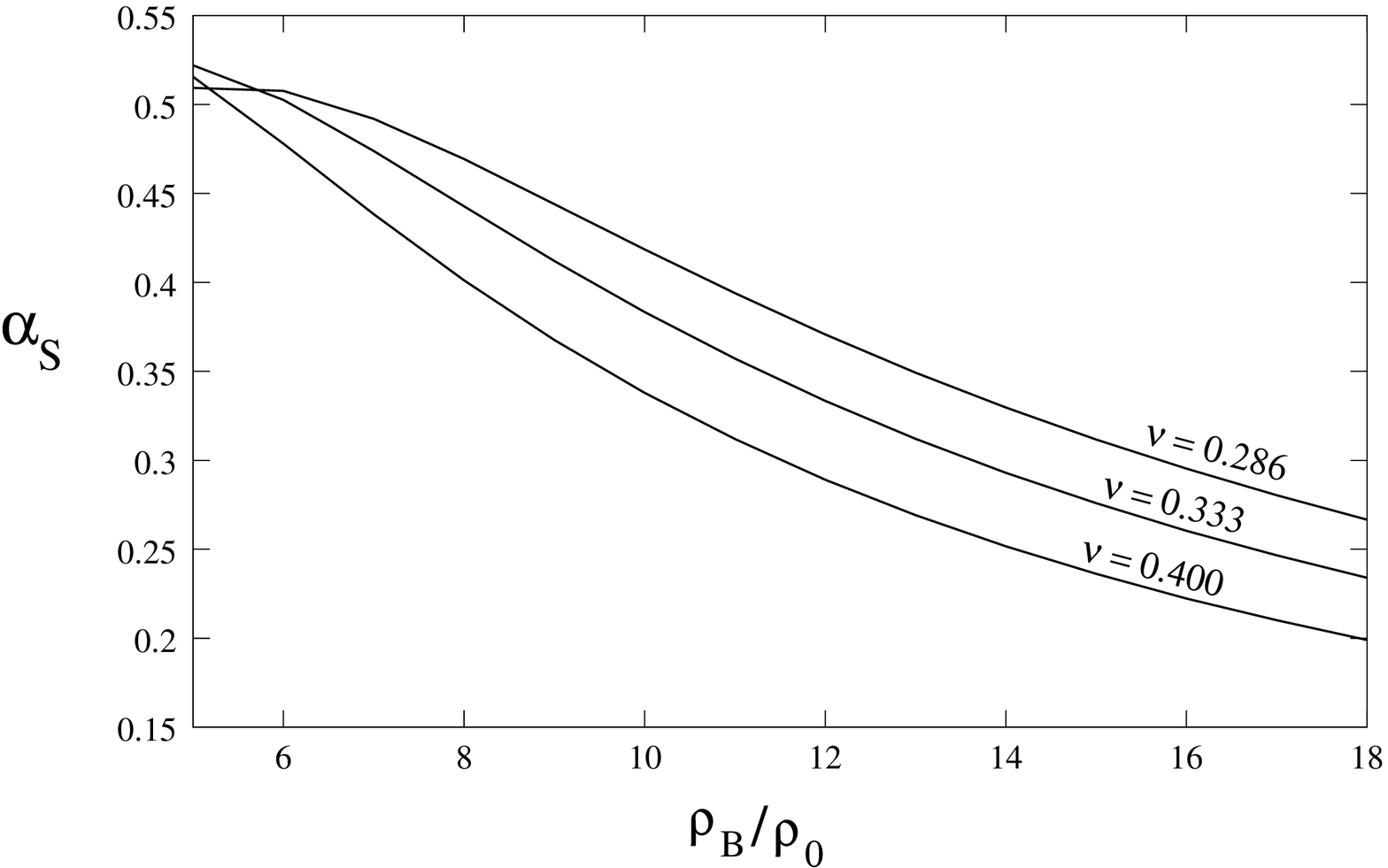,height=8.5cm}}}
\caption{The figure shows the smooth decrease in the value of the strong
coupling constant with the increase of baryonic density. We have plotted
three curves corresponding to the three values of the parameter $\nu$ ( =
0.286, 0.333 and 0.040) of eq. (\ref{eq:qm}).}
\label{fig:alph}
\end{figure}

Now we introduce the second term for finite $\rho$ but $T=0$ so that the 
Fermi function reduces to a step function. 
\begin{equation}
m(\rho) = m(0) - \f{\alpha_s}{\pi}\int_0^{E_F}\f{dE}{\sqrt{E^2 - m(\rho)^2}}. 
\label{eq:finden}
\end{equation}

	Given our form for $m(\rho)$, which is $M_Q$ of eq.(\ref{eq:qm}),
dropping the current quark mass $m_i$, $\alpha_s$ is now evaluated for all
densities. For small density its value increases but this is unnecessary for
our model.  We get very reasonable values of $\alpha_s$ for $\rho = 5 \rho_0$
upwards as can be seen in Table 1 and also in Fig(\ref {fig:alph}). The
calculation serves double purposes, it shows that for the concerned densities
our choice of the mass function, namely eq.(\ref{eq:qm}), is reasonable and
also as already stressed it gives a novel shape for the variation of
$\alpha_s$ with density not easy to obtain otherwise.

	The variation with $\nu$ is very small at the surface but deep inside
the star when the density increases the values of $\alpha_s$ goes down
substantially and the change assumes significance.  The physicality of the
results will enable us to extend our calculation now to $T ~\ne~ 0$ and apply
to early universe where temperature of order $100~ MeV$ is expected.

    In summary we have shown both in principle and in practice that the
strong coupling constant can be derived at various densities once one knows
the behaviour of the chiral symmetry restoration for the quark mass
constrained from stellar data. The practical model that we have chosen uses
an empirical form for chiral restoration at high density proposed and tested
against various star properties through the allowed mass radius regions of
the compact objects\cite{1,2,3}. We can therefore urge the astrophysics
data-analysts to pinpoint the mass and radius of compact objects like SAX
J1808.8 to a greater precision to help efforts like ours.

\vskip 1.5cm
It is a  pleasure to thank Dr. Arun Thampan for helpful discussions. 
The stimulation for the work came from a discussion with Prof. Donald 
Lynden-Bell. 

\end{document}